\documentclass[lettersize,journal]{IEEEtran}

\usepackage{amsmath,amssymb,amsfonts,amsthm}
\usepackage{comment}
\usepackage{physics}
\usepackage{glossaries}

\usepackage{hyperref}
\usepackage{lipsum} 
\usepackage{xcolor}
\usepackage{url}

\usepackage[numbers,sort&compress]{natbib} 

\usepackage{amsmath}
\usepackage{numprint} 
\usepackage{amsthm} 
\usepackage{url} 
\usepackage{textcomp} 

\usepackage{accents} 
\usepackage{amssymb}
\usepackage{amsfonts}
\usepackage{dsfont}
\usepackage[figuresright]{rotating} 
\usepackage{floatpag} 
	\rotfloatpagestyle{empty} 

\usepackage{algorithm}
\usepackage{algpseudocode}
\usepackage{longtable}
\usepackage{subfigure}

\usepackage{epstopdf} 
\usepackage{makeidx} 
\usepackage{upgreek} 
\makeindex 

\usepackage{times}
\usepackage{mathdots} 

\usepackage{bm}
\usepackage{bbm}

\theoremstyle{plain}

\def\bb0{{\mathbb{0}}}

\def\bb{{\boldsymbol{b}}}

\def\bg{{\boldsymbol{g}}}
\def\bh{{\boldsymbol{h}}}

\def\bn{{\boldsymbol{n}}}

\def\bs{{\boldsymbol{s}}}

\def\bu{{\boldsymbol{u}}}
\def\bv{{\boldsymbol{v}}}
\def\bw{{\boldsymbol{w}}}
\def\bx{{\boldsymbol{x}}}
\def\by{{\boldsymbol{y}}}

\def\b0{{\boldsymbol{0}}}

\def\bA{{\boldsymbol{A}}}

\def\bC{{\boldsymbol{C}}}
\def\bD{{\boldsymbol{D}}}

\def\bG{{\boldsymbol{G}}}
\def\bH{{\boldsymbol{H}}}
\def\bI{{\boldsymbol{I}}}

\def\bN{{\boldsymbol{N}}}

\def\bQ{{\boldsymbol{Q}}}
\def\bR{{\boldsymbol{R}}}
\def\bS{{\boldsymbol{S}}}

\def\bY{{\boldsymbol{Y}}}

\def\b{{\mathrm{b}}}

\def\r0{{\mathbf{0}}}

\def\bbE{{\mathbb{E}}}

\def\bpsi{\bm \psi}

\def\bPsi{\bm \Psi}

\def\bsf0{{\bm{\mathsf{0}}}}

\def\N0{{N_{\mathrm{0}}}}
\def\sinc{\mathrm{sinc}}

\def\bsf{{\boldsymbol{s}_\mathrm{f}}}

\newacronym{mimo}{MIMO}{multiple-input multiple-output}
\newacronym{simo}{SIMO}{single-input multiple-output}
\newacronym{siso}{SISO}{single-input single-output}
\newacronym{ris}{RIS}{reconfigurable intelligence surface}
\newacronym{bs}{BS}{base station}
\newacronym{ms}{MS}{mobile station}
\newacronym{csi}{CSI}{channel state information}
\newacronym{csir}{CSIR}{channel state information at the receiver}
\newacronym{tdd}{TDD}{time division duplexing}
\newacronym{los}{LOS}{line of sight}
\newacronym{ao}{AO}{alternative optimization}
\newacronym{se}{SE}{spectral efficiency}
\newacronym{aoa}{AoA}{angle of arrival}
\newacronym{nlos}{NLOS}{Non-Line-of-Sight}
\newacronym{sinr}{SINR}{signal-to-interference-plus-noise ratio}
\newacronym{snr}{SNR}{signal-to-noise ratio}
\newacronym{mmse}{MMSE}{minimum mean squared error}
\newacronym{mse}{MSE}{mean squared error}
\newacronym{ls}{LS}{least square}
\newacronym{awgn}{AWGN}{additive white Gaussian noise}
\newacronym{iid}{IID}{independent identically distributed}
\newacronym{pas}{PAS}{power angle spectrum}
\newacronym{pdpr}{PDPR}{pilot-to-data power ratio}
\newacronym{svd}{SVD}{singular value decomposition}

\newcommand{\forMasoud}[1]{\noindent { {{$\blacktriangleright$ 
   {\textsf{[For myself]: {\color{green}#1}}} $\blacktriangleleft$}}}}

\graphicspath{{./images/}}

\begin{document}

\bstctlcite{IEEEexample:BSTcontrol}

\title{Pilot-to-Data Power Ratio in RIS-Assisted Multiantenna Communication
\thanks{Masoud Sadeghian and Angel Lozano are with the Department of Engineering, Univ. Pompeu Fabra, 08018 Barcelona, Spain (\{masoud.sadeghian, angel.lozano\}@upf.edu). Gabor Fodor is with the School of Electrical Engineering and Computer Science, KTH Royal Institute of Technology, 10044 Stockholm, Sweden and Ericsson Research, 16480 Stockholm, Sweden. Work supported by the Horizon 2020 MSCA-ITN-METAWIRELESS Grant Agreement 956256, by MICIU under grant CEX2021001195-M, by ICREA, and by the  Swedish SSF project SAICOM, Grant No: FUS21-0004.}
}

\author{\IEEEauthorblockN{Masoud~Sadeghian},
{\it Student Member,~IEEE},
\and
\IEEEauthorblockN{Angel~Lozano},
{\it Fellow,~IEEE},
\and
\IEEEauthorblockN{Gabor~Fodor},
{\it Fellow,~IEEE}
}

\maketitle
\begin{abstract}
The optimization of the \gls{pdpr} is a recourse that helps wireless systems to acquire channel state information while minimizing the pilot overhead. 
While the optimization of the \gls{pdpr} in cellular networks has been studied extensively, 
the effect of the \gls{pdpr} in \gls{ris}-assisted networks has hardly been examined.
This paper tackles this optimization when the communication is assisted by a RIS whose phase shifts are adjusted on the basis of the statistics of the channels.
For a setting representative of a macrocellular deployment, the benefits of optimizing the PDPR are seen to be significant over a broad range of operating conditions.
These benefits, demonstrated through the ergodic minimum mean squared error, for which a closed-form solution is derived, become more pronounced as the number of RIS elements and/or the channel coherence grow large.
\end{abstract}

\smallskip
\begin{IEEEkeywords}
Multiantenna communication, reconfigurable intelligent surface, pilot power boosting 
\end{IEEEkeywords}

\glsresetall 

\section{Introduction}
\label{Chapter: Intro.}

In order to minimize the number of pilot symbol transmissions, and increase their effectiveness,
wireless systems can in principle allocate unequal powers to pilot and data symbols, subject to an average power constraint \cite{AL_PDPR,fodor2021performance}. 
Indeed, many wireless systems do feature a \gls{pdpr} that is not unity, but rather skewed towards the pilots.
The 3GPP Long-Term Evolution and New Radio systems, for example, facilitate adjusting the number of pilot symbols and boosting their power \cite{Elgendi:21}. 
This paper seeks to expand this understanding to the new paradigm in which
the communication is assisted by a \gls{ris}.


With strategic \gls{ris} placement and phase shift adjustment, a \gls{ris} can help overcome the limitations of a weak direct link, facilitating communication in scenarios where it would normally be unfeasible \cite{BasarSurvey}.
A \gls{ris} drawback, though, is its inability to process received pilot symbols for channel estimation, due to the passive nature of its elements.
Estimation of the channels to/from a \gls{ris} thus becomes extremely burdensome.
To address this issue, some works have attempted to reduce the pilot overhead by grouping the \gls{ris} elements \cite{EG_pilotred} (see also \cite{RIS_PDPR2,EB_RIS_PDPR}, which explore the grouping of elements in conjunction with \gls{pdpr} in single-input single-output \gls{ris}-assisted communication).
Nevertheless, the system still requires a prohibitively large number of pilot transmissions \cite{CSyWZ_ChEst_RIS,MDyMS_ChEst_RIS}.
An enticing alternative is to adjust the phase shifts based only on the channel statistics, rather than on its realization. This is sometimes referred to as the two-timescale scheme, in reference to the short-term adaptation of the receiver and the long-term adaptation of the \gls{ris} phase shifts \cite{BN_two-timescale_RIS,PP_two-timescale_RIS,MX_two-timescale_RIS,MDyMS_two-timescale_RIS}.
The two-timescale approach dramatically shrinks the number of symbols that must be reserved for pilot transmissions, down to the amount required for the receiver to estimate the cascade channel---exactly as in non-\gls{ris}-assisted communication.
To further minimize the number of pilot symbol transmissions and increase their effectiveness, the optimization of the \gls{pdpr} emerges as an opportunity.

This paper proposes to jointly configure the \gls{pdpr} and the \gls{ris} phase shifts such that both are based on the two-timescale operation.
This configuration problem is cast as an optimization task, with an iterative algorithm proposed in \cite{ASyPD_RISoptimization} employed for the computation of the \gls{ris} phase shifts, while an original closed-form expression is derived for the \gls{pdpr}.
With a view to representing a macrocellular deployment where the \gls{ris} is situated above the scattering clutter, the channel between the \gls{bs} and the \gls{ris} is taken to be \gls{los}, while the channel between the \gls{ris} and \gls{ms} exhibits correlated fading. 
Given this setup, the focus is on the uplink, with \gls{mmse} reception. 
We hasten to emphasize that the actual MMSE receiver is formulated, rather than the one that would take the channel estimate as exact and then minimize the signal mean-square error under that premise \cite{MinMSE:15}. The two receivers need not coincide, depending on how the channel is estimated and what its statistics are.
The \gls{ris} phase shifts are, as highlighted earlier, adjusted in a statistical manner.
Altogether, this paper contributes to the existing literature by:
\begin{itemize}
    \item Deriving a closed form for the ergodic \gls{mmse} of the actual \gls{mmse} receiver 
    in RIS-assisted communication.
    \item Establishing the optimal \gls{pdpr} as a function of the fading coherence, the number of antennas at the \gls{bs}, the number of \gls{ris} elements, and the \gls{ris} phase shifts.
\end{itemize}
Through the optimization of the \gls{pdpr} for the \gls{mmse} receiver in practical scenarios with explicit channel estimation, this paper illustrates the feasibility of improving the ergodic \gls{mmse} in \gls{ris}-assisted communication, with gains of up to $3$ dB.

\section{System Model}
\label{Chapter: Sys. Model}

Consider a single-antenna \gls{ms} communicating with an $N_\text{b}$-antenna \gls{bs}, aided by a \gls{ris} equipped with $N_\text{r}$ elements.
Assuming the channel remains constant over a fading coherence period of $\tau_\text{c}$ symbols, the \gls{ms} first transmits pilot symbols for the purpose of channel estimation, subsequently followed by payload data symbols.

\subsection{Uplink Pilot Signal Model}
\label{Chapter: Uplink pilot}

The pilot symbols conform to a Zadoff-Chu sequence of length $\tau_\text{p}$, namely
\begin{equation}
    \bs = \left[s_1, s_2, \dots,s_{\tau_\text{p}}\right]^\text{T} \in \mathbb{C}^{\tau_\text{p} \times 1},
    \label{eq:pilot sym.}
\end{equation}
whose entries satisfy $|s_j|^2=1$.
The observed pilot sequence $\bY_\text{p} \in \mathbb{C}^{ N_\text{b} \times \tau_\text{p}}$ at the \gls{bs} is
\begin{align}
    \bY_\text{p} = \alpha \sqrt{P_\text{p}} \bh \bs^\text{T} + \bN_\text{p},
    \label{eq:received pilot}
\end{align}
where $\alpha$ represents the large-scale channel gain, $P_\text{p}$ is the transmit power over the pilot sequence, and $\bN_\text{p} \in \mathbb{C}^{N_\text{b} \times \tau_\text{p}}$ is \gls{awgn}, with independent entries having variance $\sigma^2$.
The \gls{ris} becomes crucial when the direct link is weak or obstructed \cite{BasarSurvey,EB-LS-PP_RIS_magazine}. In this case, the cascade channel $\bh \in \mathbb{C}^{N_\text{b} \times 1}$ is formed by the \gls{ris} reflections, as detailed in Sec. \ref{Chapter: Channel Model}, while the direct link is assumed 
to be obstructed.
Letting $\text{vec}(\cdot)$ denote the column stacking operator,
\begin{align}
     \by_\text{p} & =  \text{vec}(\bY_\text{p}) = \alpha \sqrt{P_\text{p}} \bS \bh  + \bn_\text{p},
    \label{eq:receivedpilot_column}
\end{align}
where $\by_\text{p} \in \mathbb{C}^{\tau_\text{p} N_\text{b} \times 1} $ is the vectorized pilot sequence, $\bn_\text{p} \in \mathbb{C}^{\tau_\text{p} N_\text{b} \times 1} $ is the vectorized \gls{awgn}, and the semiorthogonal matrix $\bS = \bs \otimes \bI_{N_\text{b}} \in \mathbb{C}^{\tau_\text{p} N_\text{b} \times N_\text{b}}$ is defined by means of the Kronecker product.

\subsection{Uplink Data Signal Model}
\label{Chapter: data signal model}

Upon payload data transmission, the observation $\by_\text{d} \in \mathbb{C}^{N_\text{b} \times 1}$  at the \gls{bs} is
\begin{align}
    \by_\text{d} = \alpha \sqrt{P_\text{d}} \bh x + \bn_\text{d},
    \label{eq:received data}
\end{align}
where $\bn_\text{d} \sim \mathcal{N}_\mathbb{C}(\mathbf{0},\sigma^2\bI_{N_\text{b}})$ is the AWGN, $P_\text{d}$ is the transmit power, and $x$ is a unit-variance data symbol.

\subsection{Cascade Channel}
\label{Chapter: Channel Model}

The channel between the \gls{bs} and the \gls{ris}, represented by $\bH \in \mathbb{C}^{N_\text{b} \times N_\text{r}}$, is \gls{los}, deterministic and of unit rank.
The channel between the \gls{ms} and the \gls{ris} is $\bh_{\text{r}} \sim \mathcal{N}_\mathbb{C}(\mathbf{0},\bC_\text{r})$, with covariance matrix $\bC_\text{r} \in \mathbb{C}^{N_\text{r} \times N_\text{r}}$.
This covariance can be learned by empirically averaging signal observations.
With the configuration of the \gls{ris} specified by the phase-shift matrix $\bPsi = \text{diag}(\bpsi)$, where $\bpsi = [e^{-j\psi_1},\dots,e^{-j\psi_{N_\text{r}}}]^\text{T}$, the cascade channel at the \gls{bs} from the \gls{ms} is
\begin{equation}
    \bh = \bH \bPsi \bh_\text{r}.
    \label{eq:CPA_ComChannel_t}
\end{equation}

\section{Channel Estimation}
\label{Chapter: Channel Estimation}

Given the statistical optimization of the \gls{ris}, the cascade channel satisfies $\bh  \sim \mathcal{N}_\mathbb{C}(\mathbf{0},\bC)$ with covariance matrix
\begin{align}
    \bC =  \bH \bPsi \bC_\text{r} \bPsi^* \bH^\text{H}.
    \label{eq:ComCh_covariance}
\end{align}
By correlating the observations with the known pilot sequence, the \gls{bs} can obtain the least square estimate of the cascade channel as
\begin{align}
    \hat{\bh} & = \frac{1}{\alpha\sqrt{P_\text{p}}}\left( \bS^\text{H}\bS\right)^{-1} \bS^\text{H} \by_\text{p} \\
    & = \bh  + \frac{1}{\alpha\tau_\text{p}\sqrt{P_\text{p}}} \bS^\text{H} \bn_\text{p},
    \label{eq:LS_base}
\end{align}
satisfying $\hat{\bh}  \sim \mathcal{N}_\mathbb{C}(\mathbf{0},\bR)$ with
\begin{align}
    \bR  = \bC + \frac{\sigma^2}{ \alpha^2 \tau_\text{p} P_\text{p} } \bI_{N_\text{b}}.
    \label{eq:LS_covariance}
\end{align}


Regardless of the channel estimation method, the formulation of the \gls{mmse} receiver depends on the conditional distribution of the cascade channel relative to its estimate.
This conditional distribution satisfies \cite{GaFo_MMSEreceiver_mMIMO}
\begin{equation}
\label{eq:DQ}
    \bh|\hat{\bh} \sim \mathcal{N}_\mathbb{C}(\bD\hat{\bh},\bQ),
\end{equation}
where $\bD = \bC{\bR}^{-1}$, and $\bQ = \bC- \bC{\bR}^{-1} \bC$.

\section{MMSE Receiver}
\label{Chapter: MMSE Receiver}

By definition, the MMSE receiver
$\bw \in \mathbb{C}^{N_\text{b} \times 1} $ is given by
\begin{align} 
    \bw = \arg \min_{\bg} \; \bbE \! \left[ |\bg^\text{H} \by_\text{d} - x|^{2} \big| \hat{\bh} \right],
    \label{eq:MMSE_BS_base}
\end{align}
and the solution to this minimization is
\cite{MinMSE:15}
\begin{align} \nonumber
    \bw^\text{H} & =  \alpha \sqrt{P_\text{d}} \hat{\bh}^\text{H} \bD^\text{H} \\
    & \quad \cdot \left( \alpha^2 P_\text{d} \left( \bD \hat{\bh} \hat{\bh}^\text{H} \bD^\text{H} + \bQ \right) + \sigma^2 \bI_{ N_\text{b} } \right)^{-1},
    \label{eq:MMSE_BS}
\end{align}
while the ensuing MMSE is
\begin{align} \nonumber
    \bbE \! \left[ |\bw^\text{H} \by_\text{d} - x|^{2} \big| \hat{\bh} \right] & = \alpha^2 P_\text{d} \bw^\text{H} \left( \bD \hat{\bh} \hat{\bh}^\text{H} \bD^\text{H} + \bQ\right) \bw \\ \nonumber
    & \quad - \alpha \sqrt{P_\text{d}} \left( \bw^\text{H} \bD \hat{\bh} + \hat{\bh}^\text{H} \bD^\text{H} \bw \right) \\ 
    & \quad + \sigma^2 \bw^\text{H} \bw + 1.
    \label{eq:MSE_BS}
\end{align}
Expecting over the distribution of  $\hat{\bh}$, the ergodic MMSE emerges
(see App. \ref{app:MMSE_RX}) as
\begin{align} \hspace{-1mm}
    \bbE \! \left[ |\bw^\text{H} \by_\text{d} - x|^{2} \right]  & =
    \bbE \! \left[ \frac{1} {1+\rho}\right],
    \label{eq:MSE_T}
\end{align}
where
\begin{equation}
    \rho = \hat{\bh}^\text{H} \bD^\text{H} \left( \bQ+ \frac{\sigma^2}{\alpha^2 P_\text{d}} \bI_{ N_\text{b} } \right)^{\!-1} \!\! \bD \hat{\bh} .
    \label{Barcelona}
\end{equation}

Using $\bH= \sqrt{N_\text{r} N_\text{b}} \bu \bv^\text{H}$, where $\bu$ and $\bv$ are the steering vectors at RIS and BS, both of unit norm, the covariance matrix of the cascade channel in (\ref{eq:ComCh_covariance}) can be rewritten as
\begin{align}
    \bC 
    & = N_\text{r} N_\text{b} \bu \bv^\text{H} \bPsi \bC_\text{r} \bPsi^* \bv  \bu^\text{H} \\
    & = \zeta N_\text{r} N_\text{b} \bu \bu^\text{H},
    \label{eq:Cov_rank1}
\end{align}
where $\zeta=\bv^\text{H} \bPsi \bC_\text{r} \bPsi^* \bv$ (a positive real number) captures the effect of the RIS phase shifts. The more pronounced the correlation among RIS elements, the stronger this effect. Conversely, in the absence of spatial correlations, the phase shifts become immaterial.

As shown in App. \ref{app:rank_1_eig}, $\rho$ in (\ref{Barcelona}) is exponentially distributed with parameter
\begin{equation}
    \lambda = \frac{  \zeta^2 N_\text{r}^2 N_\text{b}^2 }
    {  \zeta N_\text{r} N_\text{b} \frac{\sigma^2}{\alpha^2 \tau_\text{p} P_\text{p}} +  \zeta N_\text{r} N_\text{b} \frac{\sigma^2}{\alpha^2 P_\text{d}} + \frac{\sigma^2}{\alpha^2 \tau_\text{p} P_\text{p}} \frac{ \sigma^2 }{ \alpha^2 P_\text{d}} },
    \label{eq:singularvalue_effective}
\end{equation}
which leads to the ergodic \gls{mmse} becoming
\begin{align}
    \bbE \! \left[ \frac{1} {1+\rho}\right] 
    & = \frac{1}{\lambda } e^{\frac{1}{\lambda }} E_1 \! \left( \frac{1}{\lambda } \right),
\label{eq:MSE_Rank1_final}
\end{align}
given the exponential integral
    $
        E_{1}(x) = \int_{x}^\infty \frac{e^{-t}}{t} dt .
    $
The parameter $\lambda$ represents the ergodic signal-to-noise ratio at the output of the MMSE receiver, including, through $\zeta$, the gains provided by the RIS phase shifts, as well as the gains from having multiple RIS elements and BS antennas.
It can be verified that (\ref{eq:MSE_Rank1_final}) is a convex function of $\lambda$, which can be minimized by optimizing over $P_\text{p}$, $P_\text{d}$, and $\zeta$.

\section{Impact of the PDPR and RIS Phase Shifts}
\label{Chapter: PDPR and RIS}

\subsection{PDPR}

The optimization over $P_\text{p}$ and $P_\text{d}$ capitalizes on the ability of wireless systems to allocate unequal powers to pilot and data symbols within a given power budget, such that 
\begin{equation}
    \frac{\tau_\text{p}}{\tau_\text{c}} P_\text{p} + \frac{\tau_\text{d}}{\tau_\text{c}} P_\text{d} = P_\text{t},
\end{equation}
where $P_\text{t}$ quantifies the average transmit power constraint on a per-symbol basis, and $\tau_\text{d}=\tau_\text{c}-\tau_\text{p}$ is the number of data symbols within the coherence interval.

Let $\gamma_\text{p} = P_{\text{p}} / P_{\text{t}}$ and $\gamma_\text{d} = P_{\text{d}} / P_{\text{t}}$ be the pilot and data power ratios relative to the average power.
As the cascade channel is rank-1, hence essentially one-dimensional, $\tau_\text{p}=1$ suffices. 
Then, $\tau_\text{d} = \tau_\text{c}-1$ and the power ratios can be seen to satisfy
\begin{align}
    \gamma_\text{d} & = \frac{\tau_\text{c}- \gamma_\text{p}}{\tau_\text{c}-1}.
\end{align}
Plugging $\gamma_\text{p}$ and $\gamma_\text{d}$ in \eqref{eq:singularvalue_effective} and letting $\text{SNR}=\alpha^2 P_\text{t}/ \sigma^2$ denote the per-antenna SNR at the receiver (prior to the MMSE processing) gives
\begin{equation}
    \lambda = \frac{ \zeta^2 N_\text{r}^2 N_\text{b}^2 \text{SNR}^2 }
    {   \frac{ \tau_\text{c}^2 + (\tau_\text{c}-2) \tau_\text{c} \gamma_\text{p} }{\gamma_\text{p}(\tau_\text{c}-\gamma_\text{p})} \, \zeta N_\text{r} N_\text{b} \text{SNR} + \frac{ \tau_\text{c}^2 (\tau_\text{c}-1)}{ \gamma_\text{p}(\tau_\text{c}-\gamma_\text{p})} }.
    \label{eq:singularvalue_eff_powerratio}
\end{equation}

As $\lambda$ is convex in $\gamma_\text{p}$, the optimum value for the latter---provided in \eqref{eq:opt_power_ratio}, atop the next page---is found by setting the derivative to zero.
\begin{figure*}[t]
\begin{equation}
    \gamma_\text{p}^\star = \frac{ \frac{\tau_\text{c}(1-\tau_\text{c})}{\zeta N_\text{r} N_\text{b}\text{SNR}} - \tau_\text{c} + \tau_\text{c} \sqrt{ \frac{(\tau_\text{c}-1)^2}{\left(\zeta N_\text{r} N_\text{b}\text{SNR}\right)^2} + \frac{(\tau_\text{c}-1)^2+(\tau_\text{c}-1)}{\zeta N_\text{r} N_\text{b}\text{SNR}} + \tau_\text{c}-1}}{ (\tau_\text{c} - 2)}
    \label{eq:opt_power_ratio}
\end{equation}
\end{figure*}
For $\text{SNR} \to 0$,
$\gamma_\text{p}^\star  \rightarrow \tau_\text{c}/2$, indicating that half of the transmit energy should be injected on the pilot while the rest is spread over the data symbols; this is consistent with standard results in non-RIS-assisted multiantenna communication \cite[Ch.~4.8.2]{ALRH_MIMObook}. It follows that the \gls{pdpr} for vanishing SNR satisfies $\gamma_\text{p}^\star / \gamma_\text{d}^\star \to \tau_{\text{c}}-1$.
Conversely, for $\text{SNR} \to \infty$,
\begin{equation}
\gamma_\text{p}^\star \rightarrow \frac{\tau_\text{c}}{1+\sqrt{\tau_\text{c}-1}},
\label{summer}
\end{equation}
whereby the \gls{pdpr} satisfies
\begin{equation}
    \frac{\gamma_\text{p}^\star}{\gamma_\text{d}^\star} \to \sqrt{\tau_\text{c}-1} .
    \label{Peke}
\end{equation}
As implied by \eqref{Peke}, the peakedness of the PDPR grows with $\sqrt{\tau_\text{c}}$ for large $\tau_\text{c}$. Only for very large $\tau_\text{c}$ could this become excessive, and in that case the PDPR could be truncated with negligible loss in performance.

\begin{figure}[t]
    \centering
    \includegraphics[width=.97\linewidth]{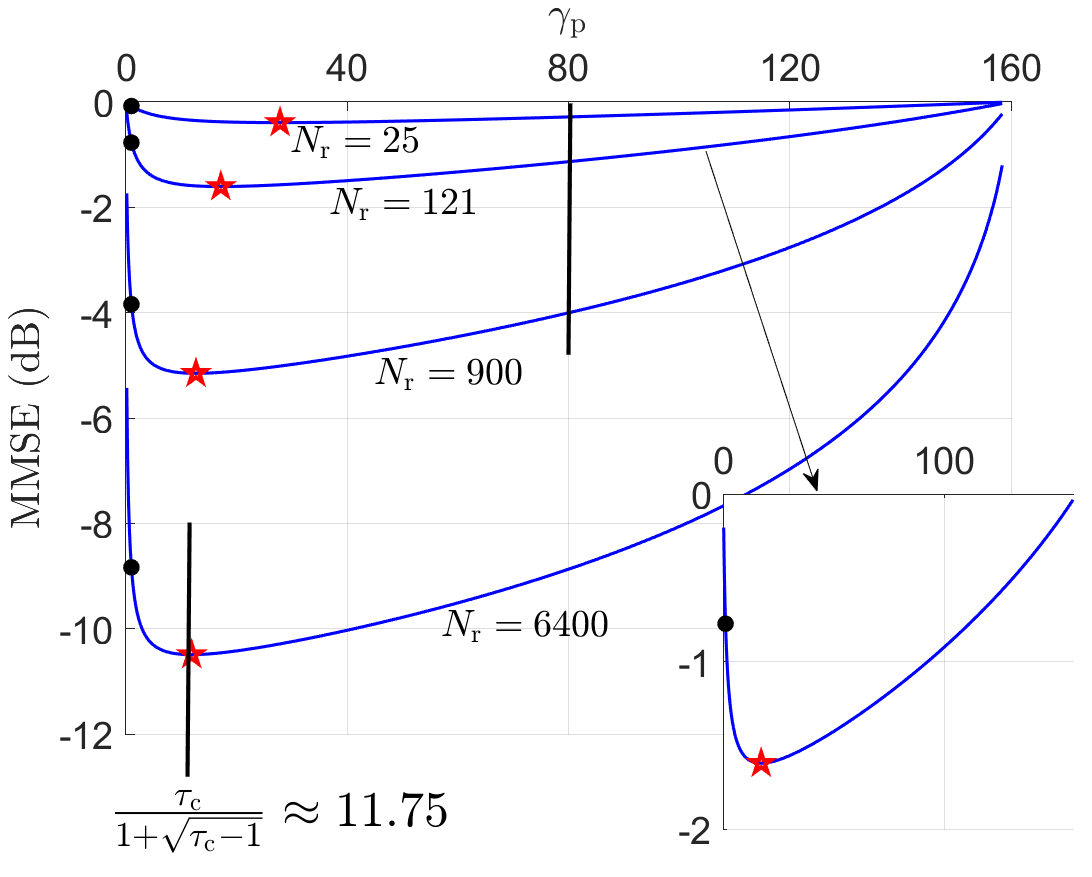} 
    \caption{Ergodic \gls{mmse} as a function of $0<\gamma_\text{p}<\tau_\text{c}$ for various numbers of RIS elements, $\tau_\text{c}=160$, and $\zeta N_\text{b}\text{SNR}=0$ dB.}
    \label{fig:opt_PDPR}
\end{figure}

Fig. \ref{fig:opt_PDPR} shows the ergodic \gls{mmse} as a function of $\gamma_\text{p}$ for different numbers of \gls{ris} elements, $N_\text{r}$, with the high-SNR limit in \eqref{summer} being approached 
as the number of RIS elements grows large for a given $\zeta N_\text{b}\text{SNR}$.
The stars indicate the optimal pilot power boosting in \eqref{eq:opt_power_ratio}, while the two vertical lines indicate its low- and high-SNR limits.
The solid circles indicate the ergodic \gls{mmse} when equal power is allocated to pilot and data symbols, i.e., for $\gamma_\text{p}=1$.
As the number of RIS elements increases, the gap between the stars and the solid circles widens, indicating a higher advantage from optimizing the PDPR.
\begin{comment}
\begin{figure}[t]
    \centering
    \includegraphics[width=.95\linewidth]{Figure/MMSE_basedongammap_V5_blackcircle.eps} \vspace{-1.6mm}
    \caption{\textcolor{blue}{Ergodic \gls{mmse} in dB over the interval $0<\gamma_\text{p}<\tau_\text{c}$ with varying numbers of RIS elements, $\tau_\text{c}=160$, and $\zeta N_\text{b}\text{SNR}=0$ dB. The red stars indicate the optimal pilot power boosting in \eqref{eq:opt_power_ratio}, while approaching $\gamma_\text{p}\approx 11.75$ in the high-SNR limits. The filled black circles represent the case where equal power is allocated to pilot and data symbols, $\gamma_\text{p}=1$.}}
    \label{fig:opt_PDPR}
    \vspace{-1.6mm}
\end{figure}
\end{comment}

\subsection{RIS Phase Shifts}

Since $\lambda$ is strictly increasing in $\zeta$, due to $\zeta$ being positive and real, the optimum phase shifts are those that maximize $\zeta$.
Rewriting it as $\zeta=  \bpsi^\text{H} \bG \bpsi$, where $\bG = \text{diag}(\bv^\text{H}) \bC_\text{r} \text{diag}(\bv)$, the optimization of the RIS phase shifts can be cast as
\begin{align}
    (\text{P}1): \quad 
    & \max_{\bpsi} \quad \bpsi^\text{H} \bG \bpsi \\
    & \; \text{s.t.} \quad \quad |\psi_i| = 1 \qquad i = 1, 2, \dots, N_\text{r},
\label{eq:RIS_opt}
\end{align}
whose objective function is quadratic, but whose unit-magnitude constraints are not convex. 
The optimization problem $(\text{P}1)$ was investigated in \cite{ASyPD_RISoptimization,NByJFW_RISoptimization}, where dimension-wise sinusoidal maximization (DSM) was identified as an effective solution. This paper applies a modified DSM approach \cite{ASyPD_RISoptimization}. The phase shift of the $n$th RIS element, $\psi_n^{(k)}$, is optimized, while keeping the phases of all other elements fixed, leading to
\begin{equation}
    \psi_n^{(k)}=\exp \! \left(\angle\left[\sum_{m=1}^{n-1} \psi_m^{(k)} g_{nm} + \sum_{m=n+1}^{N_\text{r}} \psi_m^{(k-1)} g_{nm} \right]\right),
    \label{eq:DSM}
\end{equation}
where $g_{nm} = [\bG]_{nm}$.
Algorithm $1$ outlines the implementation, starting with an initial $\bpsi^{(0)}$.
\begin{algorithm} [t]
\caption{Computation of \gls{ris} phase shifts via DSM}
\begin{algorithmic}[1] 
\State \textbf{input:} $\bG$ and $\epsilon$
\State \textbf{output:} $\bpsi^{(k)}$ and $\zeta^{(k)}={\bpsi^{(k)}}^{\text{H}} \bG \bpsi^{(k)}$
\State \textbf{initialize} $k=0$ and set an initial $\bpsi^{(0)}$
\State \textbf{compute} $g_{nm}=[\bG]_{nm}$
\While{$|\zeta^{(k)} - \zeta^{(k-1)}| > \epsilon$}
    \State increase $k$
        \For{$n=1,\dots, N_\text{r}$}
            \State Update $\psi_n^{(k)}$ using \eqref{eq:DSM}
        \EndFor
\EndWhile
\end{algorithmic}
\label{alg:AO_RIS}
\end{algorithm}

\begin{figure}[t]
    \centering
    \includegraphics[width=.97\linewidth]{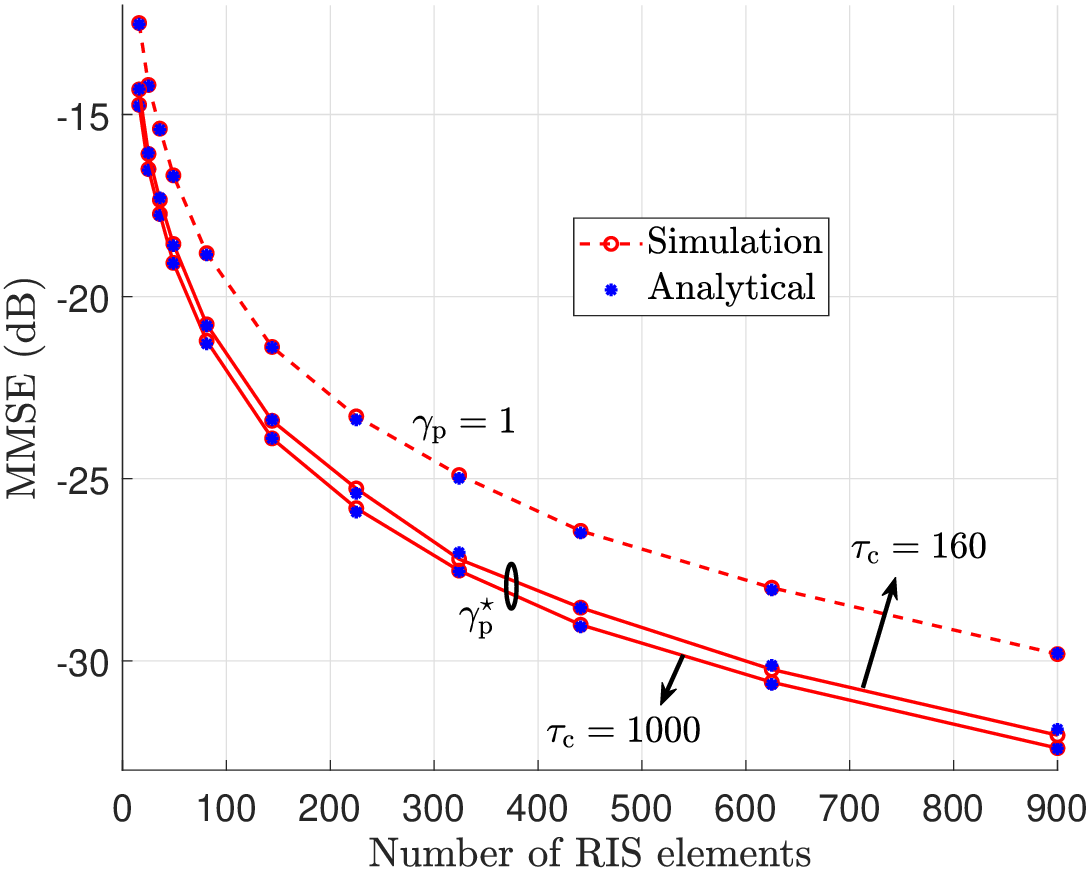} \vspace{-1.6mm}
    \caption{Ergodic \gls{mmse} versus number of \gls{ris} elements for two distinct coherence intervals.
    The dashed line corresponds to $\gamma_{\text{p}}=1$ (any fading coherence) while the solid lines correspond to the optimal pilot power boosting in \eqref{eq:opt_power_ratio}. 
    Algorithm~1 is used to obtain the \gls{ris} phase shifts for both the dashed and solid lines.
    }
    \label{fig:MSE_comparison}
\end{figure}

In an isotropic scattering environment, the most disfavorable one for a RIS with statistical phase shifting, the $(n,m)$th entry of $\bC_{\text{r}}$ equals \cite{EmBjLuSa_RFadingChModel_RIS}
\begin{equation}
    c_{n,m} = \sinc\! \left( \frac{2 \, \|\bu_n-\bu_m\|}{\lambda_\text{c}}\right) \qquad n,m = 1, \dots, N_\text{r} ,
    \label{eq:RIS_PAS_halfspace}
\end{equation}
where $\bu_n$ and $\bu_m$ are the positions of the $n$th and $m$th \gls{ris} elements, respectively, and $\lambda_\text{c}$ is the carrier wavelength.
Considering a planar square RIS and a BS with $N_{\text{b}}=10$ antennas arranged as a $2 \times 5$ planar array with half-wavelength spacing, Fig. \ref{fig:MSE_comparison} depicts the ergodic \gls{mmse} as a function of the number of RIS elements.
The spacing between the \gls{ris} elements is $\lambda_\text{c}/7$ along each dimension, and SNR$=-10$~dB. 
When the number of \gls{ris} elements is small, increasing it causes $\lambda$ in
\eqref{eq:singularvalue_eff_powerratio} to grow quadratically, which leads to a steeper reduction in ergodic \gls{mmse}. Conversely, with a large number of \gls{ris} elements, the increase in $\lambda$ becomes linear, causing the ergodic \gls{mmse} to decrease more slowly as additional \gls{ris} elements are introduced.

\section{Conclusion}
\label{Chapter: Conclusion} 
This paper has shown that a \gls{pdpr} optimization, even with a suboptimal \gls{ris} phase shift configuration such as the one embodied by Algorithm $1$, leads to a substantial performance improvement.
This advantage increases steadily with the number of \gls{ris} elements, as the system is pushed into an effectively higher SNR.
The advantage also increases noticeably with the fading coherence,
but, even for short coherences, gains of roughly $3$ dB can be observed over a broad range of operating conditions of interest for macrocellular systems.

\appendices

\section{} \label{app:MMSE_RX}
From \eqref{eq:MSE_BS}, the ergodic \gls{mmse} is
\begin{align}
    \label{eq:MSE_T_appendix} 
    \bbE \! \left[ |\bw^\text{H} \by_\text{d} - x|^{2} \right] & =
   \bbE_{\hat{\bh}} \Bigg[ 1 -  2 \underbrace{ \alpha \sqrt{P_\text{d}} \bw^\text{H} \bD \hat{\bh} }_{\text{term $1$}} \\  \nonumber
    & \!\!\!\!\!\!\!\!\!\! +  \underbrace{\alpha^2 P_\text{d} \bw^\text{H} \left( \bD \hat{\bh} \hat{\bh}^\text{H} \bD^\text{H} + \bQ + \frac{\sigma^2}{\alpha^2 P_\text{d}} \bI_{ N_\text{b} }\right) \bw}_{\text{term $2$}} \Bigg].
\end{align}
Recalling \eqref{eq:MMSE_BS}, letting $\bA = \bQ+ \frac{\sigma^2}{\alpha^2 P_\text{d}} \bI_{ N_\text{b} }$, and applying the Sherman–Morrison formula, term $1$ emerges as
\begin{align} \label{eq:term1_A1}
    & \!\!\!\! \hat{\bh}^\text{H} \bD^\text{H} \left( \bD \hat{\bh} \hat{\bh}^\text{H} \bD^\text{H} + \bA \right)^{-1} \bD \hat{\bh} \nonumber \\
    & \qquad = \hat{\bh}^\text{H} \bD^\text{H} \left( \bA^{-1} - \frac{ \bA^{-1} \bD \hat{\bh} \hat{\bh}^\text{H} \bD^\text{H} \bA^{-1} }{ 1 + \hat{\bh}^\text{H} \bD^\text{H} \bA^{-1} \bD \hat{\bh} } \right) \bD \hat{\bh} \\
    & \qquad = \frac{ \hat{\bh}^\text{H} \bD^\text{H} \bA^{-1}  \bD \hat{\bh} }{ 1 + \hat{\bh}^\text{H} \bD^\text{H} \bA^{-1} \bD \hat{\bh} }
    \label{eq:term1_A}
\end{align}
while term $2$ becomes
\begin{align} \nonumber
    & \!\!\!\!\!\!\!\!\!\!\!\!\!\! \hat{\bh}^\text{H} \bD^\text{H} \left( \bD \hat{\bh} \hat{\bh}^\text{H} \bD^\text{H} + \bA \right)^{-1} \left( \bD \hat{\bh} \hat{\bh}^\text{H} \bD^\text{H} + \bA \right) \\ 
    & \!\!\!\!\!\!\!\!\!\!\!\!\!\! \nonumber \cdot \left( \bD \hat{\bh} \hat{\bh}^\text{H} \bD^\text{H} + \bA \right)^{-1} \bD \hat{\bh} \\
    & \qquad\quad = \hat{\bh}^\text{H} \bD^\text{H} \left( \bD \hat{\bh} \hat{\bh}^\text{H} \bD^\text{H} + \bA \right)^{-1} \bD \hat{\bh} \\
    & \qquad\quad = \frac{\hat{\bh}^\text{H} \bD^\text{H} \bA^{-1} \bD \hat{\bh}}{1+\hat{\bh}^\text{H} \bD^\text{H} \bA^{-1} \bD \hat{\bh}},
    \label{eq:term2_base}
\end{align}
which mirrors term $1$, whose symmetry simplifies \eqref{eq:MSE_T_appendix} to
\begin{align}
    \bbE \! \left[ |\bw^\text{H} \by_\text{d} - x|^{2} \right]
    & = \bbE \! \left[ \frac{1} {1+\rho}\right],
    \label{eq:MSE_BS2}
\end{align}
with $\rho=\hat{\bh}^\text{H} \bD^\text{H} \bA^{-1} \bD \hat{\bh}$.

\section{} \label{app:rank_1_eig}
By plugging \eqref{eq:Cov_rank1} and simplifying,
\begin{align}
    \bR^{-1} & =\left( \frac{\sigma^2}{ \alpha^2 \tau_\text{p} P_\text{p}} \bI_{ N_\text{b} } +  \zeta N_\text{r} N_\text{b} \bu \bu^\text{H} \right)^{-1} \\
    \label{eq:SMf_forR}
    & = \frac{1}{ \zeta N_\text{r} N_\text{b}} \frac{  \zeta N_\text{r} N_\text{b} \alpha^2 \tau_\text{p} P_\text{p}}{\sigma^2} \left( \bI_{ N_\text{b} } - \frac{\bu \bu^\text{H}}{1+\frac{\sigma^2}{  \zeta N_\text{r} N_\text{b} \alpha^2 \tau_\text{p} P_\text{p}}} \right) \\
    & = \frac{ \alpha^2 \tau_\text{p} P_\text{p}}{\sigma^2} \left( \bI_{ N_\text{b} } - \frac{  \zeta N_\text{r} N_\text{b} \alpha^2 \tau_\text{p} P_\text{p} }{  \zeta N_\text{r} N_\text{b} \alpha^2 \tau_\text{p} P_\text{p}+\sigma^2 }\bu \bu^\text{H}\right),
\end{align}
where \eqref{eq:SMf_forR} follows from the Sherman–Morrison formula. Thus,
\begin{align}
    \bC \bR^{-1} \bC & = \frac{   \zeta^2 N_\text{r}^2 N_\text{b}^2 \alpha^2 \tau_\text{p} P_\text{p}}{\sigma^2} \bu \bu^\text{H} \Bigg( \bI_{ N_\text{b} } \nonumber \\ 
    & \quad - \frac{  \zeta N_\text{r} N_\text{b} \alpha^2 \tau_\text{p} P_\text{p} }{  \zeta N_\text{r} N_\text{b} \alpha^2 \tau_\text{p} P_\text{p}+\sigma^2 }\bu \bu^\text{H}\Bigg) \bu \bu^\text{H} \\
    & = \frac{ \zeta^2 N_\text{r}^2 N_\text{b}^2 }{ \zeta N_\text{r} N_\text{b} +\frac{\sigma^2}{\alpha^2 \tau_\text{p} P_\text{p}}} \bu \bu^\text{H},
\end{align}
which leads to
\begin{align}
\bQ & = \bC- \bC \bR^{-1} \bC \\ 
&  = \frac{ \zeta N_\text{r} N_\text{b} \frac{\sigma^2}{\alpha^2 \tau_\text{p} P_\text{p}}}{ \zeta N_\text{r} N_\text{b} + \frac{\sigma^2}{\alpha^2 \tau_\text{p} P_\text{p}}} \bu \bu^\text{H}.
\label{eq:MMSE_error_rank1}
\end{align}
After some linear algebra, $\bD {\bR}^{1/2} = (\bC \bR^{-1} \bC )^{1/2}$, giving 
\begin{align}
    \bD {\bR}^{1/2}
    = & \sqrt{ \frac{ \zeta^2 N_\text{r}^2 N_\text{b}^2 }{ \zeta N_\text{r} N_\text{b} +\frac{\sigma^2}{\alpha^2 \tau_\text{p} P_\text{p}}} } \bu \bu^\text{H}.
\label{eq:MMSE_mid2_rank1}
\end{align}
Applying the Sherman–Morrison formula again, 
\begin{align} \nonumber
& \left( \bQ + \frac{\sigma^2}{\alpha^2 P_\text{d}} \bI_{ N_\text{b} } \right)^{-1} \\
& \;\; = \left( \frac{ \zeta N_\text{r} N_\text{b} \frac{\sigma^2}{\alpha^2 \tau_\text{p} P_\text{p}}}{ \zeta N_\text{r} N_\text{b} + \frac{\sigma^2}{\alpha^2 \tau_\text{p} P_\text{p}}} \bu \bu^\text{H} + \frac{\sigma^2}{\alpha^2 P_\text{d}} \bI_{ N_\text{b} } \right)^{-1} \\ 
& \;\; = \frac{\alpha^2 P_\text{d}}{\sigma^2} \Bigg( \bI_{ N_\text{b} } \nonumber \\
& \quad\quad - \frac{  \zeta N_\text{r} N_\text{b} \frac{\sigma^2}{\alpha^2 \tau_\text{p} P_\text{p}} }{  \zeta N_\text{r} N_\text{b} \frac{\sigma^2}{\alpha^2 \tau_\text{p} P_\text{p}} +  \zeta N_\text{r} N_\text{b} \frac{\sigma^2}{\alpha^2 P_\text{d}} + \frac{\sigma^2}{\alpha^2 \tau_\text{p} P_\text{p}} \frac{\sigma^2}{\alpha^2 P_\text{d}}}  \bu \bu^\text{H}  \Bigg),
\label{eq:MMSE_mid1_rank1}
\end{align}
and altogether
\begin{align} \nonumber
    & \bR^{1/2} \bD^\text{H} \left( \bQ + \frac{\sigma^2} {\alpha^2 P_\text{d}} \bI_{ N_\text{b} } \right)^{-1} \bD \bR^{1/2} \\ \nonumber
    & \;\; = \frac{   \zeta^2 N_\text{r}^2 N_\text{b}^2 }{ \zeta N_\text{r} N_\text{b} +\frac{\sigma^2}{\alpha^2 \tau_\text{p} P_\text{p}}} \frac{\alpha^2 P_\text{d}}{\sigma^2} \Bigg( \bu \bu^\text{H} \\
    & \;\;\;\;\;\; - \frac{  \zeta N_\text{r} N_\text{b} \frac{\sigma^2}{\alpha^2 \tau_\text{p} P_\text{p}} }{  \zeta N_\text{r} N_\text{b} \frac{\sigma^2}{\alpha^2 \tau_\text{p} P_\text{p}} +  \zeta N_\text{r} N_\text{b} \frac{\sigma^2}{\alpha^2 P_\text{d}} + \frac{\sigma^2}{\alpha^2 \tau_\text{p} P_\text{p}} \frac{ \sigma^2 }{ \alpha^2 P_\text{d}} } \bu \bu^\text{H}  \Bigg) \nonumber \\ 
    & \;\; = \underbrace{\frac{  \zeta^2 N_\text{r}^2 N_\text{b}^2 }
    { \zeta N_\text{r} N_\text{b} \frac{\sigma^2}{\alpha^2 \tau_\text{p} P_\text{p}} +  \zeta N_\text{r} N_\text{b} \frac{\sigma^2}{\alpha^2 P_\text{d}} + \frac{\sigma^2}{\alpha^2 \tau_\text{p} P_\text{p}} \frac{ \sigma^2 }{ \alpha^2 P_\text{d}} }}_{\lambda} \bu \bu^\text{H}.
    \label{eq:MMSE_mid3_rank1}
\end{align}
Plugged into (\ref{Barcelona}), with $\hat{\bh} = \bR^{1/2} \bh_\text{u}$ where $\bh_\text{u} \sim \mathcal{N}_\mathbb{C}(\mathbf{0},\bI_{ N_\text{b} })$,
the above evidences that $\rho$ is exponentially distributed with parameter $\lambda$.

{
\small
\bibliographystyle{IEEEtran}
\bibliography{refs}

\begin{thebibliography}{10}
\providecommand{\url}[1]{#1}
\csname url@samestyle\endcsname
\providecommand{\newblock}{\relax}
\providecommand{\bibinfo}[2]{#2}
\providecommand{\BIBentrySTDinterwordspacing}{\spaceskip=0pt\relax}
\providecommand{\BIBentryALTinterwordstretchfactor}{4}
\providecommand{\BIBentryALTinterwordspacing}{\spaceskip=\fontdimen2\font plus
\BIBentryALTinterwordstretchfactor\fontdimen3\font minus
  \fontdimen4\font\relax}
\providecommand{\BIBforeignlanguage}[2]{{%
\expandafter\ifx\csname l@#1\endcsname\relax
\typeout{** WARNING: IEEEtran.bst: No hyphenation pattern has been}%
\typeout{** loaded for the language `#1'. Using the pattern for}%
\typeout{** the default language instead.}%
\else
\language=\csname l@#1\endcsname
\fi
#2}}
\providecommand{\BIBdecl}{\relax}
\BIBdecl

\bibitem{AL_PDPR}
A.~Lozano, ``Interplay of spectral efficiency, power and {Doppler} spectrum for
  reference-signal-assisted wireless communication,'' \emph{IEEE Tran. Wireless
  Comm.}, vol.~7, no.~12, pp. 5020--29, 2008.

\bibitem{fodor2021performance}
G.~Fodor \emph{et~al.}, ``Performance analysis of a linear {MMSE} receiver in
  time-variant {Rayleigh} fading channels,'' \emph{IEEE Trans. Commun.},
  vol.~69, no.~6, pp. 4098--4112, 2021.

\bibitem{Elgendi:21}
H.~Elgendi \emph{et~al.}, ``Uplink performance of {LTE} and {NR} with
  high-speed trains,'' in \emph{IEEE Veh. Techn. Conf. (VTC Spring)}, 2021.

\bibitem{BasarSurvey}
E.~Basar \emph{et~al.}, ``{Wireless communications through reconfigurable
  intelligent surfaces},'' \emph{IEEE Access}, vol.~7, pp. 116\,753--116\,773,
  2019.

\bibitem{EG_pilotred}
Y.~Yang \emph{et~al.}, ``Intelligent reflecting surface meets {OFDM}: Protocol
  design and rate maximization,'' \emph{IEEE Trans. Commun.}, vol.~68, no.~7,
  pp. 4522--4535, 2020.

\bibitem{RIS_PDPR2}
Y.~Li \emph{et~al.}, ``Joint power allocation and passive beamforming for
  {RIS}-assisted wireless energy-constrained systems with multi-user
  scheduling,'' \emph{IEEE Trans. Veh. Technol.}, vol.~72, no.~11, pp.
  15\,109--15\,114, 2023.

\bibitem{EB_RIS_PDPR}
A.~Enqvist \emph{et~al.}, ``Optimizing reconfigurable intelligent surfaces for
  short transmissions: How detailed configurations can be afforded?''
  \emph{IEEE Trans. Wireless Commun.}, vol.~23, no.~4, pp. 3377--3391, 2024.

\bibitem{CSyWZ_ChEst_RIS}
Z.~Wang \emph{et~al.}, ``{Channel estimation for intelligent reflecting surface
  assisted multiuser communications: Framework, algorithms, and analysis},''
  \emph{IEEE Trans. Wireless Commun.}, vol.~19, no.~10, pp. 6607--6620, 2020.

\bibitem{MDyMS_ChEst_RIS}
H.~Alwazani \emph{et~al.}, ``{Intelligent reflecting surface-assisted
  multi-user MISO communication: Channel estimation and beamforming design},''
  \emph{IEEE Open J. Commun. Soc.}, vol.~1, pp. 661--680, 2020.

\bibitem{BN_two-timescale_RIS}
K.~Zhi \emph{et~al.}, ``{Two-timescale design for reconfigurable intelligent
  surface-aided massive MIMO systems with imperfect CSI},'' \emph{IEEE Trans.
  Inf. Theory}, vol.~69, no.~5, pp. 3001--3033, 2022.

\bibitem{PP_two-timescale_RIS}
Q.~Peng \emph{et~al.}, ``Two-timescale design for reconfigurable intelligent
  surface-aided {URLLC},'' \emph{IEEE Trans. Wireless Commun.}, vol.~23,
  no.~10, pp. 13\,664--13\,677, 2024.

\bibitem{MX_two-timescale_RIS}
Y.~Han \emph{et~al.}, ``{Large intelligent surface-assisted wireless
  communication exploiting statistical CSI},'' \emph{IEEE Trans. Veh.
  Technol.}, vol.~68, no.~8, pp. 8238--8242, 2019.

\bibitem{MDyMS_two-timescale_RIS}
A.~Kammoun \emph{et~al.}, ``{Asymptotic max-min SINR analysis of reconfigurable
  intelligent surface assisted MISO systems},'' \emph{IEEE Trans. Wireless
  Commun.}, vol.~19, no.~12, pp. 7748--7764, 2020.

\bibitem{ASyPD_RISoptimization}
A.~Sirojuddin \emph{et~al.}, ``Low-complexity sum-capacity maximization for
  intelligent reflecting surface-aided {MIMO} systems,'' \emph{IEEE Wireless
  Commun. Lett.}, vol.~11, no.~7, pp. 1354--1358, 2022.

\bibitem{MinMSE:15}
G.~Fodor \emph{et~al.}, ``On minimizing the {MSE} in the presence of channel
  state information errors,'' \emph{IEEE Commun. Lett.}, vol.~19, no.~9, pp.
  1604--1607, 2015.

\bibitem{EB-LS-PP_RIS_magazine}
E.~Björnson \emph{et~al.}, ``Reconfigurable intelligent surfaces: A signal
  processing perspective with wireless applications,'' \emph{IEEE Signal
  Process. Mag.}, vol.~39, no.~2, pp. 135--158, 2022.

\bibitem{GaFo_MMSEreceiver_mMIMO}
G.~Fodor \emph{et~al.}, ``{Performance analysis of block and comb type channel
  estimation for massive {MIMO} systems},'' in \emph{Proc. 1st Int. Conf. on
  5G}, 2014, pp. 62--69.

\bibitem{ALRH_MIMObook}
R.~W. Heath~Jr and A.~Lozano, \emph{Foundations of MIMO Communication}.\hskip
  1em plus 0.5em minus 0.4em\relax Cambridge, U.K.: Cambridge Univ. Press,
  2019.

\bibitem{NByJFW_RISoptimization}
B.~Ning \emph{et~al.}, ``Beamforming optimization for intelligent reflecting
  surface assisted {MIMO}: A sum-path-gain maximization approach,'' \emph{IEEE
  Wireless Commun. Lett.}, vol.~9, no.~7, pp. 1105--1109, 2020.

\bibitem{EmBjLuSa_RFadingChModel_RIS}
E.~Björnson and L.~Sanguinetti, ``Rayleigh fading modeling and channel
  hardening for reconfigurable intelligent surfaces,'' \emph{IEEE Wireless
  Commun. Lett.}, vol.~10, no.~4, pp. 830--834, 2021.

\end{thebibliography}
}
\end{document}